\documentclass[a4paper,draftcls,onecolumn,technote,11pt]{IEEEtran}%
\usepackage{amsfonts}
\usepackage{amsmath}
\usepackage{amssymb}
\usepackage{graphicx}%
\providecommand{\U}[1]{\protect\rule{.1in}{.1in}}
\begin{document}

\title{Controllability Analysis for Multirotor Helicopter Rotor Degradation and Failure}
\author{Guang-Xun Du, Quan Quan, Binxian Yang, Kai-Yuan Cai\thanks{The authors are
with Department of Automatic Control, Beihang University, Beijing 100191,
China (dgx@asee.buaa.edu.cn; qq\_buaa@buaa.edu.cn;
yangbinxian@asee.buaa.edu.cn; kycai@buaa.edu.cn)}}
\maketitle

\section*{Nomenclature}%

\begin{tabular}
[c]{@{}lcl}%
$h$ & = & altitude of the helicopter, m\\
$\phi,\theta,\psi$ & = & roll, pitch and yaw angles of the helicopter, rad\\
$v_{h}$ & = & vertical velocity of the helicopter, m/s\\
$p,q,r$ & = & roll, pitch and yaw angular velocities of the helicopter,
rad/s\\
$T$ & = & total thrust of the helicopter, N\\
$L,M,N$ & = & airframe roll, pitch and yaw torque of the helicopter, N$\cdot
$m\\
$m_{a}$ & = & mass of the helicopter, kg\\
$g$ & = & acceleration of gravity, kg$\cdot$m/s$^{2}$\\
$J_{x},J_{y},J_{z}$ & = & moment of inertia around the roll, pitch and yaw
axes of the\\
&  & helicopter frame, kg$\cdot$m$^{2}$\\
$f_{i}$ & = & lift of the $i$-th rotor, N\\
$K_{i}$ & = & maximum lift of the $i$-th rotor, N\\
$\eta_{i}$ & = & efficiency parameter of the $i$-th rotor\\
$r_{i}$ & = & distance from the center of the $i$-th rotor to the center of
mass, m\\
$m$ & = & number of rotors\\
$k_{\mu}$ & = & ratio between the reactive torque and the lift of the rotors
\end{tabular}

\section{Introduction}

Multirotor helicopters \cite{Robotics,ASM-Mechatronic,Bill(2011)} are
attracting increasing attention in recent years because of their important
contribution and cost effective application in several tasks such as
surveillance, search and rescue missions and so on. However, there exists a
potential risk to civil safety if a mutirotor aircraft crashes, especially in
an urban area. Therefore, it is of great importance to consider the flight
safety of multirotor helicopters in the presence of rotor faults or failures
\cite{Iman}.

Fault-Tolerant Control (FTC) \cite{FTFC} has the potential to improve the
safety and reliability of multirotor helicopters. FTC is the ability of a
controlled system to maintain or gracefully degrade control objectives despite
the occurrence of a fault \cite{Yang}. There are many applications in which
fault tolerance may be achieved by using adaptive control, reliable
control\textbf{,} or reconfigurable control strategies \cite{YMZ-JJ,AFTFC}.
Some strategies involve explicit fault diagnosis, and some do not. The reader
is referred to a recent survey paper \cite{Bib-YMZ} for an outline of the
state of art in the field of FTC. However, only few attempts are known that
focus on the fundamental FTC property analysis, one of which is defined as the
(control) reconfigurability \cite{Yang}. A faulty multirotor system with
inadequate reconfigurability cannot be made to effectively tolerate faults
regardless of the feedback control strategy used \cite{NEWu}. The control
reconfigurability can be analyzed from the intrinsic and performance-based
perspectives. The aim of this Note is to analyze the control reconfigurability
for multirotor systems (4-, 6- and 8-rotor helicopters, etc.) from the
controllability analysis point of view\textbf{.}

Classical controllability theories of linear systems are not sufficient to
test the controllability of the considered multirotor helicopters, as the
rotors can only provide unidirectional lift (upward or downward) in practice.
In our previous work \cite{Du-JIRS}, it was shown that a hexacopter with the
standard symmetrical configuration is uncontrollable if one rotor fails,
though the controllability matrix of the hexacopter is row full rank. Thus,
the reconfigurability based on the controllability Gramian \cite{NEWu} is no
longer applicable. Brammer in \cite{Brammer(1972)} proposed a necessary and
sufficient condition for the controllability of linear autonomous systems with
positive constraint, which can be used to analyze the controllability of
multirotor systems. However, the theorems in \cite{Brammer(1972)} are not easy
to use in practice. Owing to this, the controllability of a given system is
reduced to those of its subsystems with real eigenvalues based on the Jordan
canonical form in \cite{Hiroshi Yoshida}. However, appropriate stable
algorithms to compute Jordan real canonical form should be used to avoid
ill-conditioned calculations. Moreover, a step-by-step controllability test
procedure is not given. To address these problems, in this Note the theory
proposed in \cite{Brammer(1972)} is extended and a new necessary and
sufficient condition of controllability is derived for the considered
multirotor systems.

Nowadays, larger multirotor aircraft are starting to emerge and some
multirotor aircraft are controlled by varying the collective pitch of the
blade. This work considers only the multirotor helicopters controlled by
varying the RPM (Revolutions Per Minute) of each rotor but this research can
be extended to most multirotor aircraft regardless of size whether they are
controlled by varying the collective pitch of the blade or the RPM.

The linear dynamical model of the considered multirotor helicopters around
hover conditions is derived first, and then the control constraint is
specified. It is pointed out that classical controllability theories of linear
systems are not sufficient to test the controllability of the derived model
(Section II). Then the controllability of the derived model is studied based
on the theory in \cite{Brammer(1972)}, and two conditions which are necessary
and sufficient for the controllability of the derived model are given. In
order to make the two conditions easy to test in practice, an Available
Control Authority Index (ACAI) is introduced to quantify the available control
authority of the considered multirotor systems. Based on the ACAI, a new
necessary and sufficient condition is given to test the controllability of the
considered multirotor systems (Section III). Furthermore, the computation of
the proposed ACAI and a step-by-step controllability test procedure is
approached for practical application (Section IV). The proposed
controllability test method is used to analyze the controllability of a class
of hexacopters to show its effectiveness (Section V). The major contributions
of this Note are: (i) an ACAI to quantify the available control authority of
the considered multirotor systems, (ii) a new necessary and sufficient
controllability test condition based on the proposed ACAI, and (iii) a
step-by-step controllability test procedure for the considered multirotor systems.

\section{Problem Formulation}

\begin{figure}[ptb]
\begin{center}
\includegraphics[
scale=0.7 ]{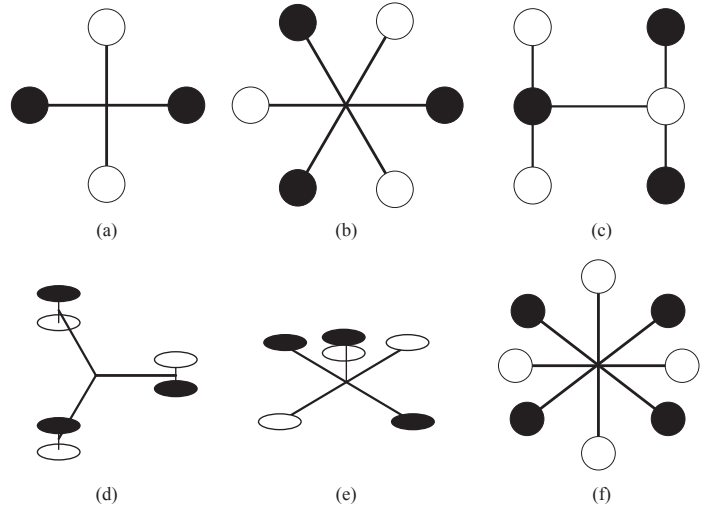}
\end{center}
\caption{Different configurations of multirotor helicopters (the white
disc\ denotes that the rotor rotates clockwise and the black disc denotes that
the rotor rotates anticlockwise)}%
\label{multirotors}%
\end{figure}

This Note considers a class of multirotor helicopters shown in
Fig.\ref{multirotors}, which are often used in practice. From
Fig.\ref{multirotors}, it can be seen that there are various types of
multirotor helicopters with different rotor numbers and different
configurations. Despite the difference in type and configuration, they can all
be modeled in a general form as equation (\ref{linear}). In reality, the
dynamical model of the multirotor helicopters is nonlinear and there are some
aerodynamic damping and stiffness. But if the multirotor helicopter is
hovering, the aerodynamic damping and stiffness is ignorable. The linear
dynamical model around hover conditions is given as
\cite{Guillaume(2011),Du-ASDB,imav2012}:
\begin{equation}
\dot{x}=Ax+B\underset{u}{\underbrace{\left(  F-G\right)  }} \label{linear}%
\end{equation}
where%
\begin{align*}
x  &  =\left[  h\text{ }\phi\text{ }\theta\text{ }\psi\text{ }v_{h}\text{
}p\text{ }q\text{ }r\right]  ^{T}\in%
\mathbb{R}
^{8},F=\left[  T\text{ }L\text{ }M\text{ }N\right]  ^{T}\in%
\mathbb{R}
^{4},G=\left[  m_{a}g\text{ }0\text{ }0\text{ }0\right]  ^{T}\in%
\mathbb{R}
^{4},\\
A  &  =%
\begin{bmatrix}
0_{4\times4} & I_{4}\\
0 & 0
\end{bmatrix}
\in%
\mathbb{R}
^{8\times8},B=%
\begin{bmatrix}
0\\
J_{f}^{-1}%
\end{bmatrix}
\in%
\mathbb{R}
^{8\times4},J_{f}=\text{diag}\left(  -m_{a},J_{x},J_{y},J_{z}\right)
\end{align*}

\begin{figure}[ptb]
\begin{center}
\includegraphics[
scale=0.8 ]{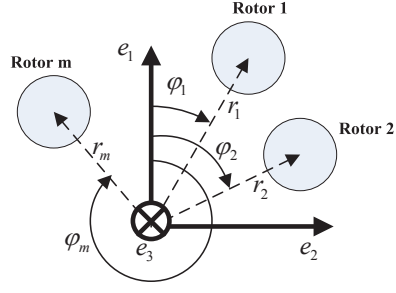}
\end{center}
\caption{Geometry definition for multirotor system}%
\label{Geometry_mutirotor}%
\end{figure}

In practice, $f_{i}\in\left[  0,K_{i}\right]  ,i=1,\cdots m$ since the rotors
can only provide unidirectional lift (upward or downward). As a result, the
rotor lift $f$ is constrained by
\begin{equation}
f\in\mathcal{F}=\Pi_{i=1}^{m}\left[  0,K_{i}\right]  . \label{f_constrain}%
\end{equation}
Then according to the geometry of the multirotor system shown in
Fig.\ref{Geometry_mutirotor}, the mapping from the rotor lift $f_{i}%
,i=1,\cdots m$ to the system total thrust/torque $F$ is:%
\begin{equation}
F=B_{f}f \label{F=Bff}%
\end{equation}
where $f=\left[  f_{1}\text{ }\cdots\text{ }f_{m}\right]  ^{T}$. The matrix
$B_{f}\in%
\mathbb{R}
^{4\times m}$ is the control effectiveness matrix and%
\begin{equation}
B_{f}=\left[  b_{1}\text{ }b_{2}\text{ }\cdots\text{ }b_{m}\right]  \label{Bf}%
\end{equation}
where $b_{i}=\eta_{i}\bar{b}_{i}$, $\bar{b}_{i}\in%
\mathbb{R}
^{4},i\in\left\{  1,\cdots m\right\}  $ is the vector of contribution factors
of the $i$-th rotor to the total thrust/torque $F$, the parameters $\eta
_{i}\in\left[  0,1\right]  ,i=1,\cdots,6$ is used to account for rotor
wear/failure. If the $i$-th rotor fails, then $\eta_{i}=0$. For a multirotor
helicopter whose geometry is shown in Fig.\ref{Geometry_mutirotor}, the
control effectiveness matrix $B_{f}$ in parameterized form is \cite{imav2012}%
\begin{equation}
B_{f}=\left[
\begin{array}
[c]{ccc}%
\eta_{1} & \cdots & \eta_{m}\\
-\eta_{1}r_{1}\sin\left(  \varphi_{1}\right)  & \cdots & -\eta_{m}r_{m}%
\sin\left(  \varphi_{m}\right) \\
\eta_{1}r_{1}\cos\left(  \varphi_{1}\right)  & \cdots & \eta_{m}r_{m}%
\cos\left(  \varphi_{m}\right) \\
\eta_{1}w_{1}k_{\mu} & \cdots & \eta_{m}w_{m}k_{\mu}%
\end{array}
\right]  \label{Bf-example}%
\end{equation}
where $w_{i}$ is defined by%
\begin{equation}
w_{i}=\left\{
\begin{array}
[c]{l}%
\text{ }1\text{, if rotor }i\text{ rotates anticlockwise}\\
-1\text{, if rotor }i\text{ rotates clockwise}%
\end{array}
\right.  . \label{si}%
\end{equation}

By (\ref{f_constrain}) and (\ref{F=Bff}), $F$ is constrained by%
\begin{equation}
\Omega=\left\{  F|F=B_{f}f,f\in\mathcal{F}\right\}  . \label{Omega}%
\end{equation}
Then $u$ is constrained by%
\begin{equation}
\mathcal{U}=\left\{  u|u=F-G,F\in\Omega\right\}  . \label{U_constrain}%
\end{equation}
From (\ref{f_constrain}) (\ref{Omega}) and (\ref{U_constrain}), $\mathcal{F}%
,\Omega,\mathcal{U},$ are all convex and closed.

Our major objective is to study the controllability of the system
(\ref{linear}) under the constraint $\mathcal{U}$.

\textbf{Remark 1.} The system (\ref{linear}) with constraint set
$\mathcal{U}\subset%
\mathbb{R}
^{4}$ is called controllable if, for each pair of points $x_{0}\in%
\mathbb{R}
^{8}$ and $x_{1}\in%
\mathbb{R}
^{8}$, there exists a bounded admissible control, $u\left(  t\right)
\in\mathcal{U}$, defined on some finite interval $0\leq t\leq t_{1}$, which
steers $x_{0}$ to $x_{1}$. Specifically, the solution to (\ref{linear}),
$x\left(  t,u\left(  \cdot\right)  \right)  $, satisfies the boundary
conditions $x\left(  0,u\left(  \cdot\right)  \right)  =x_{0}$ and $x\left(
t_{1},u\left(  \cdot\right)  \right)  =x_{1}$.

\textbf{Remark 2}. Classical controllability theories of linear systems often
require the origin to be an interior point of $\mathcal{U}$\ so that
$\mathcal{C}\left(  A,B\right)  \ $being row full rank is a necessary and
sufficient condition \cite{Brammer(1972)}. However, the origin is not always
inside control constraint $\mathcal{U}$ of the system (1) under rotor
failures. Consequently, $\mathcal{C}\left(  A,B\right)  $ being row full rank
is not sufficient to test the controllability of the system (\ref{linear}).

\section{Controllability for the Multirotor Systems}

In this section, the controllability of the system (\ref{linear}) is studied
based on the positive controllability theory proposed in \cite{Brammer(1972)}.
Applying the positive controllability theorem in \cite{Brammer(1972)} to the
system (\ref{linear}) directly, the following theorem is obtained

\textbf{Theorem 1}. The following conditions are necessary and sufficient for
the controllability of the system (\ref{linear}):

\begin{enumerate}
\item[(i)] Rank $\mathcal{C}\left(  A,B\right)  =8$, where $\mathcal{C}\left(
A,B\right)  =\left[  B\text{ }AB\text{ }\cdots\text{ }A^{7}B\right]  .$

\item[(ii)] There is no real eigenvector $v$ of $A^{T}$ satisfying
$v^{T}Bu\leq0$ for all $u\in\mathcal{U}.$
\end{enumerate}

It is difficult to test the condition (ii) in \emph{Theorem 1}, because in
practice one cannot check all $u$ in $\mathcal{U}$. In the following, an
easy-to-use criterion is proposed to test the condition (ii) in \emph{Theorem
1}. Before going further, a measure is defined as:%
\begin{equation}
\rho\left(  X,\partial\Omega\right)  \triangleq\left\{
\begin{array}
[c]{c}%
\min\left\{  \left\Vert X-F\right\Vert :X\in\Omega,F\in\partial\Omega\right\}
\\
-\min\left\{  \left\Vert X-F\right\Vert :X\in\Omega^{C},F\in\partial
\Omega\right\}
\end{array}
\right.  \label{Rou}%
\end{equation}
where $\partial\Omega$ is the boundary of $\Omega$ and $\Omega^{C}$ is the
complementary set of $\Omega$. If $\rho\left(  X,\partial\Omega\right)  \leq
0$, then $X\in\Omega^{C}\cup\partial\Omega$, which means that $X$ is not an
interior point of $\Omega$. Otherwise, $X$ is an interior point of $\Omega$.

According to (\ref{Rou}), $\rho\left(  G,\partial\Omega\right)  =\min\left\{
\left\Vert G-F\right\Vert ,F\in\partial\Omega\right\}  $ which is the radius
of the biggest enclosed sphere centered at $G$ in the attainable control set
$\Omega$. In practice, it is the maximum control thrust/torque that can be
produced in all directions. Therefore, it is an important quantity to ensure
controllability for arbitrary rotor wear/failure. Then $\rho\left(
G,\partial\Omega\right)  $ can be used to quantify the available control
authority of the system (\ref{linear}). From (\ref{U_constrain}), it can be
seen that all the elements in $\mathcal{U}$ are given by translating the all
the elements in $\Omega$ by a constant $G$. As translation does not change the
relative position of all the elements of $\Omega$, the value of $\rho\left(
0,\partial\mathcal{U}\right)  $ is equal to the value of $\rho\left(
G,\partial\Omega\right)  $. In this Note, the Available Control Authority
Index (ACAI) of system (\ref{linear}) is defined by $\rho\left(
G,\partial\Omega\right)  $ as $\Omega$ is the attainable control set and more
intuitive than $\mathcal{U}$ in practice. The ACAI shows the ability as well
as the control capacity of a multirotor helicopter controlling its altitude
and attitude. With this definition, the following lemma about condition (ii)
of \emph{Theorem 1} is obtained.

\textbf{Lemma 1:} The following three statements are equivalent for the system
(\ref{linear}):

\begin{enumerate}
\item[(i)] There is no non-zero real eigenvector $v$ of $A^{T}$ satisfying
$v^{T}Bu\leq0$ for all $u\in\mathcal{U}$ or $v^{T}B\left(  F-G\right)  \leq0$
for all $F\in\Omega$.

\item[(ii)] $G$ is an interior point of $\Omega$.

\item[(iii)] $\rho\left(  G,\partial\Omega\right)  >0$.
\end{enumerate}

\textit{Proof:} See \emph{Appendix A}. $\square$

By \emph{Lemma 1}, condition (ii) in \emph{Theorem 1} can be tested by the
value $\rho\left(  G,\partial\Omega\right)  $. Now a new necessary and
sufficient condition can be derived to test the controllability of the system
(\ref{linear}).

\textbf{Theorem 2:} System (\ref{linear}) is controllable, if and only if the
following two conditions hold:

\begin{enumerate}
\item[(i)] Rank $\mathcal{C}\left(  A,B\right)  =8$.

\item[(ii)] $\rho\left(  G,\partial\Omega\right)  >0$.
\end{enumerate}

According to \emph{Lemma 1}, \emph{Theorem 2} is straightforward from
\emph{Theorem 1}. Actually, \emph{Theorem 2} is a corollary of \emph{Theorem
1.4} presented in \cite{Brammer(1972)}. To make this Note more readable and
self-contained, we extend the condition (1.6) of \emph{Theorem 1.4} presented
in \cite{Brammer(1972)}, and get the condition (ii) in \emph{Theorem 2} of
this Note based on the simplified structure of ($A,B$) pair and the convexity
of $\mathcal{U}$. This extension can enable the quantification of the
controllability and also make it possible to develop a step-by-step
controllability test procedure for the multirotor systems. In the following
section, a step-by-step controllability test procedure is approached based on
\emph{Theorem 2}.

\section{A Step-by-Step Controllability Test Procedure}

This section will show how to obtain the value of the proposed ACAI in Section
III. Furthermore, a step-by-step controllability test procedure for the
controllability of the system (\ref{linear}) is approached for practical applications.

\subsection{Available Control Authority Index Computation}

First, two index matrices $S_{1}$ and $S_{2}$ are defined, where $S_{1}$ is a
matrix whose rows consist of all possible combinations of $3$ elements of
$M=[1$ $2$ $\cdots$ $m]$, and the corresponding rows of $S_{2}$ are the
remaining $m-3$ elements of $M$. The matrix $S_{1}$ contains $s_{m}$ rows and
$3$ columns, and the matrix $S_{2}$ contains $s_{m}$ rows and $m-3$ columns,
where%
\begin{equation}
s_{m}=\frac{m!}{\left(  m-\left(  n_{\Omega}-1\right)  \right)  !\left(
n_{\Omega}-1\right)  !}. \label{Sm}%
\end{equation}
For the system in equation\textbf{ }(\ref{linear}), $s_{m}$ is the number of
the groups of parallel boundary segments in $\mathcal{F}$. For example, if
$m=4$, $n_{\Omega}=4$, then $s_{m}=4$ and
\[
S_{1}=\left[
\begin{array}
[c]{ccc}%
1 & 2 & 3\\
1 & 2 & 4\\
1 & 3 & 4\\
2 & 3 & 4
\end{array}
\right]  ,S_{2}=\left[
\begin{array}
[c]{c}%
4\\
3\\
2\\
1
\end{array}
\right]
\]
Define $B_{1,j}$ and $B_{2,j}$ as follows:
\begin{align}
B_{1,j}  &  =[b_{S_{1}\left(  j,1\right)  }\text{ }b_{S_{1}\left(  j,2\right)
}\text{ }b_{S_{1}\left(  j,3\right)  }]\in%
\mathbb{R}
^{4\times3}\nonumber\\
B_{2,j}  &  =[b_{S_{2}\left(  j,1\right)  }\text{ }\cdots\text{ }%
b_{S_{2}\left(  j,m-3\right)  }]\in%
\mathbb{R}
^{4\times\left(  m-3\right)  } \label{Bj1Bj2}%
\end{align}
where $j=1,\cdots,s_{m}$, $S_{1}\left(  j,k_{1}\right)  $ is the element at
the $j$-th row and the $k_{1}$-th column of $S_{1}$, and $S_{2}\left(
j,k_{2}\right)  $ is the element at the $j$-th row and the $k_{2}$-th column
of $S_{2}$. Here $k_{1}=1,2,3$ and $k_{2}=1,\cdots,m-3$.

Define a sign function sign$\left(  \cdot\right)  $ as follows: for an $n$
dimensional vector $a=[a_{1}$ $\cdots$ $a_{n}]\in%
\mathbb{R}
^{1\times n}$,
\begin{equation}
\text{sign}\left(  a\right)  =[c_{1}\text{ }\cdots\text{ }c_{n}]
\label{sign_function}%
\end{equation}
where $c_{i}=1$ if $a_{i}>0$, $c_{i}=0$ if $a_{i}=0$, and $c_{i}=-1$ if
$a_{i}<0$. Then $\rho\left(  G,\partial\Omega\right)  $ is obtained by the
following theorem.

\textbf{Theorem 3}. For the system in equation (\ref{linear}), if rank
$B_{f}=4$ then the ACAI $\rho\left(  G,\partial\Omega\right)  $ is given by%
\begin{equation}
\rho\left(  G,\partial\Omega\right)  =\text{sign}\left(  \min\left(
d_{1},d_{2},\cdots,d_{s_{m}}\right)  \right)  \min\left(  \left\vert
d_{1}\right\vert ,\left\vert d_{2}\right\vert ,\cdots,\left\vert d_{s_{m}%
}\right\vert \right)  . \label{RouG}%
\end{equation}
If rank $B_{1,j}=3$, then%
\begin{equation}
d_{j}=\frac{1}{2}\text{sign}\left(  \xi_{j}^{T}B_{2,j}\right)  \Lambda
_{j}\left(  \xi_{j}^{T}B_{2,j}\right)  ^{T}-\left\vert \xi_{j}^{T}\left(
B_{f}f_{c}-G\right)  \right\vert ,j=1,\cdots,s_{m} \label{dj}%
\end{equation}
where $f_{c}=\frac{1}{2}[K_{1}$ $K_{2}\ \cdots K_{m}]^{T}\in%
\mathbb{R}
^{m}$ and $\Lambda_{j}\in%
\mathbb{R}
^{(m-3)\times(m-3)}$ is given by%
\begin{equation}
\Lambda_{j}=\left[
\begin{array}
[c]{cccc}%
K_{S_{2}\left(  j,1\right)  } & 0 & 0 & 0\\
0 & K_{S_{2}\left(  j,2\right)  } & 0 & 0\\
0 & 0 & \ddots & 0\\
0 & 0 & 0 & K_{S_{2}\left(  j,m-3\right)  }%
\end{array}
\right]  \label{Lanmda}%
\end{equation}
The vector $\xi_{j}\in%
\mathbb{R}
^{4}$ satisfies%
\begin{equation}
\xi_{j}^{T}B_{1,j}=0,\left\Vert \xi_{j}\right\Vert =1 \label{kexiB1=0}%
\end{equation}
and $B_{1,j}$ and $B_{2,j}$ are given by (\ref{Bj1Bj2})\textbf{. }If rank
$B_{1,j}<3$, $d_{j}=+\infty$.

\emph{Proof:} The proof process is divided into 3 steps and the details can be
found in \emph{Appendix B}. $\square$

\textbf{Remark 3}. In practice, $+\infty$ is replaced by a sufficiently large
positive number (for example, set $d_{j}=10^{6}$). If rank $B_{f}<4$, then
$\Omega$ is not a 4 dimensional hypercube and the ACAI makes no sense which is
set to $-\infty$. Similarly, $-\infty$ is replaced by $-10^{6}$ in practice).
From (\ref{RouG}), if $\rho\left(  G,\partial\Omega\right)  >0$, then $G$ is
an interior point of $\Omega$ and $\rho\left(  G,\partial\Omega\right)  $ is
the minimum distance from $G$ to $\partial\Omega$. If $\rho\left(
G,\partial\Omega\right)  <0$, then$\ G$ is not an interior point of $\Omega$
and $\left\vert \rho\left(  G,\partial\Omega\right)  \right\vert $ is the
minimum distance from $G$ to $\partial\Omega$. The ACAI $\rho\left(
G,\partial\Omega\right)  $ can also be used to show a degree of
controllability (see \cite{Degree,DC-Definition,DisRejection}) of the system
in equation (\ref{linear}), but the ACAI is fundamentally different from the
degree of controllability in \cite{Degree}. The degree of controllability\ in
\cite{Degree} is defined based on the minimum Euclidean norm of the state on
the boundary of the recovery region for time $t$. However, the ACAI is defined
based on the minimum Euclidean norm of the control force on the boundary of
the attainable control set. The degree of controllability in \cite{Degree}\ is
time-dependent, whereas the ACAI is time-independent. A very similar
multirotor failure assessment was provided in\emph{ }\cite{imav2012} by
computing the radius of the biggest circle that fits in the $L$-$M$ plane with
the center in the origin ($L=0$, $M=0$), where the $L$-$M$ plane is obtained
by cuting the four-dimensional attainable control set at the nominal hovering
conditions defined with $T=G$ and $N=0$. This computation is very simple and
intuitive. But the radius of the two-dimensional $L$-$M$ plane can only
quantify the control authority of roll and pitch control. To account for this,
the ACAI proposed by this Note is defined by the radius of the biggest ball
that fits in the four-dimensional polytopes $\Omega$ with the center in $G$.

\subsection{Controllability Test Procedure for Multirotor Systems}

From the above, the controllability of the multirotor system (\ref{linear})
can be analyzed by the following procedure:

\emph{Step 1:} Check the rank of $\mathcal{C}\left(  A,B\right)  $. If
$\mathcal{C}\left(  A,B\right)  =8$, go to \emph{Step 2}. If $\mathcal{C}%
\left(  A,B\right)  <8$, go to \emph{Step 9.}

\emph{Step 2}: Set the value of the rotor's efficiency parameter $\eta_{i}%
$,$i=1,\cdots,m$ to get $B_{f}=\left[  b_{1}\text{ }b_{2}\text{ }\cdots\text{
}b_{m}\right]  $ as shown in (\ref{Bf}). If rank $B_{f}=4$, go to \emph{Step
3}. If rank $B_{f}<4$, let $\rho\left(  G,\partial\Omega\right)  =-10^{6}$ and
go to \emph{Step 9}.

\emph{Step 3:} Compute the two index matrices $S_{1}$ and $S_{2}$, where
$S_{1}$ is a matrix whose rows consist of all possible combinations of the $m$
elements of $M$ taken 3 at a time and the rows of $S_{2}$ are the remaining
$\left(  m-3\right)  $ elements of $M$, $M=[1$ $2$ $\cdots$ $m]$.

\emph{Step 4}: $j=1$.

\emph{Step 5}: Compute the two matrices $B_{1,j}$ and $B_{2,j}$ according to
(\ref{Bj1Bj2}).

\emph{Step 6}: If rank $B_{1,j}=3$, compute $d_{j}$ according to (\ref{dj}).
If rank $B_{1,j}<3$, set $d_{j}=10^{6}$.

\emph{Step 7}: $j=j+1$. If $j\leq s_{m}$, go to \emph{Step 5}. If $j>s_{m}$,
go to \emph{Step 8}.

\emph{Step 8:} Compute $\rho\left(  G,\partial\Omega\right)  $ according to
(\ref{RouG}).

\emph{Step 9}: If $\mathcal{C}\left(  A,B\right)  <8$ or $\rho\left(
G,\partial\Omega\right)  \leq0$, the system (\ref{linear}) is uncontrollable.
Otherwise, the system in equation (\ref{linear}) is controllable.

\section{Controllability Analysis for a Class of Hexacopters}

\begin{figure}[ptb]
\begin{center}
\includegraphics[
scale=0.5 ] {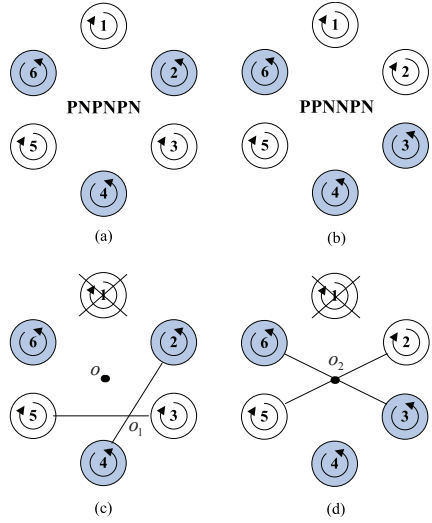}
\end{center}
\caption{(a) Standard rotor arrangement, (b) new rotor arrangement, (c) the
1-st rotor of the PNPNPN system fails, (d) the 1-st rotor of the PPNNPN system
fails.}%
\label{rotor_arrangement_total}%
\end{figure}

In this section, the controllability test procedure developed in section IV is
used to analyze the controllability of a class of hexacopters shown in
Fig.\ref{rotor_arrangement_total}, subject to rotor wear/failures, to show its effectiveness.

The rotor arrangement of the considered hexacopter is the standard symmetrical
configuration shown in Fig.\ref{rotor_arrangement_total}(a). PNPNPN is used to
denote the standard arrangement, where \textquotedblleft P\textquotedblright%
\ denotes that a rotor rotates clockwise and \textquotedblleft
N\textquotedblright\ denotes that a rotor rotates anticlockwise. According to
(\ref{Bf}), the control effectiveness matrix $B_{f}$ of that hexacopter
configuration is%
\begin{equation}
B_{f}=\left[
\begin{array}
[c]{cccccc}%
\eta_{1} & \eta_{2} & \eta_{3} & \eta_{4} & \eta_{5} & \eta_{6}\\
0 & -\frac{\sqrt{3}}{2}\eta_{2}r_{2} & -\frac{\sqrt{3}}{2}\eta_{3}r_{3} & 0 &
\frac{\sqrt{3}}{2}\eta_{5}r_{5} & \frac{\sqrt{3}}{2}\eta_{6}r_{6}\\
\eta_{1}r_{1} & \frac{1}{2}\eta_{2}r_{2} & -\frac{1}{2}\eta_{3}r_{3} &
-\eta_{4}r_{4} & -\frac{1}{2}\eta_{5}r_{5} & \frac{1}{2}\eta_{6}r_{6}\\
-\eta_{1}k_{\mu} & \eta_{2}k_{\mu} & -\eta_{3}k_{\mu} & \eta_{4}k_{\mu} &
-\eta_{5}k_{\mu} & \eta_{6}k_{\mu}%
\end{array}
\right]  \label{Bf_hexa}%
\end{equation}

\begin{table}[ptb]
\caption{Hexacopter parameters}%
\label{parameters}
\centering%
\begin{tabular}
[c]{lcc}\hline\hline
Parameter & Value & Units\\\hline
$m_{a}$ & 1.535 & kg\\
g & 9.80 & m/s$^{2}$\\
$r_{i},i=1,\cdots,6$ & 0.275 & m\\
$K_{i},i=1,\cdots,6$ & 6.125 & N\\
$J_{x}$ & 0.0411 & kg$\cdot$m$^{2}$\\
$J_{y}$ & 0.0478 & kg$\cdot$m$^{2}$\\
$J_{z}$ & 0.0599 & kg$\cdot$m$^{2}$\\
$k_{\mu}$ & 0.1 & -\\\hline\hline
\end{tabular}
\end{table}

\begin{table}[ptb]
\caption{Hexacopter (PNPNPN) controllability with one rotor failed}%
\label{controllability_one}
\centering%
\begin{tabular}
[c]{lccc}\hline\hline
Rotor failure & Rank of $\mathcal{C}(A,B)$ & ACAI & Controllability\\\hline
No wear/failure & 8 & 1.4861 & controllable\\
$\eta_{1}=0$ & 8 & 0 & uncontrollable\\
$\eta_{2}=0$ & 8 & 0 & uncontrollable\\
$\eta_{3}=0$ & 8 & 0 & uncontrollable\\
$\eta_{4}=0$ & 8 & 0 & uncontrollable\\
$\eta_{5}=0$ & 8 & 0 & uncontrollable\\
$\eta_{6}=0$ & 8 & 0 & uncontrollable\\\hline\hline
\end{tabular}
\end{table}

Using the procedure defined in Section IV, the controllability analysis
results of the PNPNPN hexacopter subject to one rotor failure is shown in
Table \ref{controllability_one}. The PNPNPN hexacopter is uncontrollable when
one rotor fails, even though its controllability matrix is row full rank. A
new rotor arrangement (PPNNPN) of the hexacopter shown in
Fig.\ref{rotor_arrangement_total}(b) is proposed in \cite{imav2012}, which is
still controllable when one of some specific rotors stops. The controllability
of the PPNNPN hexacopter subject to one rotor failure is shown in Table
\ref{controllability_one_new}.

\begin{table}[tbh]
\caption{Hexacopter (PPNNPN) controllability with one rotor failed}%
\label{controllability_one_new}
\centering%
\begin{tabular}
[c]{lccc}\hline\hline
Rotor failure & Rank of $\mathcal{C}(A,B)$ & ACAI & Controllability\\\hline
No wear/failure & 8 & 1.1295 & controllable\\
$\eta_{1}=0$ & 8 & 0.7221 & controllable\\
$\eta_{2}=0$ & 8 & 0.4510 & controllable\\
$\eta_{3}=0$ & 8 & 0.4510 & controllable\\
$\eta_{4}=0$ & 8 & 0.7221 & controllable\\
$\eta_{5}=0$ & 8 & 0 & uncontrollable\\
$\eta_{6}=0$ & 8 & 0 & uncontrollable\\\hline\hline
\end{tabular}
\end{table}

From Table \ref{controllability_one} and Table \ref{controllability_one_new},
the value of the ACAI is 1.4861 for the PNPNPN hexacopter subject to no rotor
failures, while the value of the ACAI is reduced to 1.1295 for the PPNNPN
hexacopter. It can be observed that the use of the PPNNPN configuration
instead of the PNPNPN configuration improves the fault-tolerance capabilities
but also decreases the ACAI for the no failure condition. Similar to the
results in \cite{imav2012}, changing the rotor arrangement is always a
tradeoff between fault-tolerance and control authority. That said, the PPNNPN
system is not always controllable under a failure. From Table
\ref{controllability_one_new}, it can be seen that if the 5-th rotor or the
6-th rotor fails the PPNNPN system is uncontrollable.

The following provides some physical insight between the two configurations.
For the PPNNPN configuration, if one of the rotors (other than the 5-th and
6-th rotor) of that system fails, the remaining rotors still comprise a basic
quadrotor configuration that is symmetric about the mass center (see
Fig.\ref{rotor_arrangement_total}(d)). In contrast, if one rotor of the PNPNPN
system fails, although the remaining rotors can make up a basic quadrotor
configuration, the quadrotor configuration is not symmetric about the mass
center (see Fig.\ref{rotor_arrangement_total}(c)). The result is that the
PPNNPN system under most single rotor failures can provide the necessary
thrust and torque control, while the PNPNPN system cannot.

\begin{figure}[ptb]
\begin{center}
\includegraphics[
scale=0.5 ]{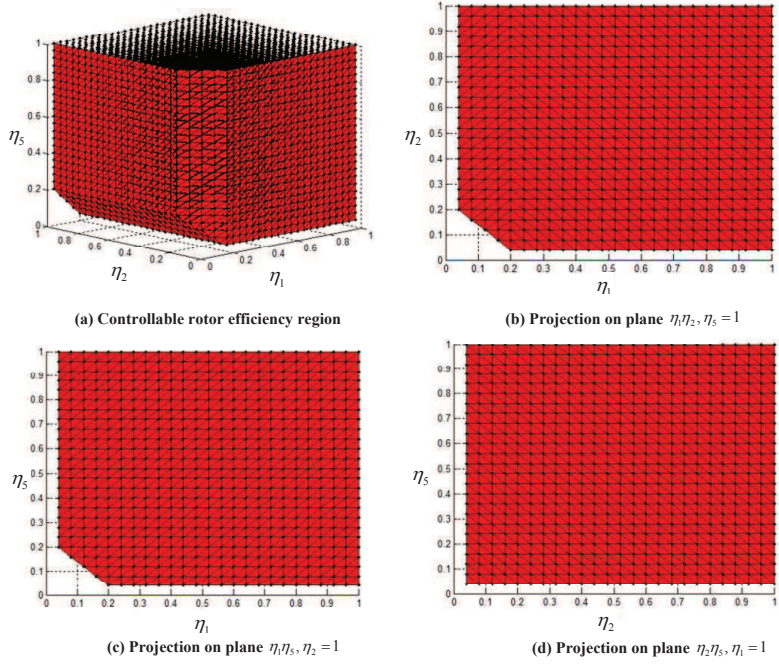}
\end{center}
\caption{Controllable region of different rotors' efficiency parameter for the
PNPNPN hexacopter}%
\label{Controllable_region}%
\end{figure}

Therefore, it is necessary to test the controllability of the multirotor
helicopters before any fault-tolerant control strategies are employed.
Moreover, the controllability test procedure approached can also be used to
test the controllability of the hexacopter with different $\eta_{i}$,
$i\in\left\{  1,\cdots,6\right\}  $. Let $\eta_{1}$, $\eta_{2}$, $\eta_{5}$
vary in $\left[  0,1\right]  \subset%
\mathbb{R}
$, namely rotor 1, rotor 2 and rotor 5 are worn; then the PNPNPN hexacopter
retains controllability while $\eta_{1}$, $\eta_{2}$, $\eta_{5}$ are in the
grid region (where the grid spacing is 0.04) in Fig.\ref{Controllable_region}.
The corresponding ACAI at the boundaries of the projections shown in Fig. 4 is
zero or near to zero (because of error in numerical calculation).

\section{Conclusions}

The controllability problem of a class of multirotor helicopters was
investigated. An Available Control Authority Index (ACAI) was introduced to
quantify the available control authority of multirotor systems. Based on the
ACAI, a new necessary and sufficient condition was given based on a positive
controllability theory. Moreover, a step-by-step procedure was developed to
test the controllability of the considered multirotor helicopters. The
proposed controllability test method was used to analyze the controllability
of a class of hexacopters to show its effectiveness. Analysis results showed
that the hexacopters with different rotor configurations have different fault
tolerant capabilities. It is therefore necessary to test the controllability
of the multirotor helicopters before any fault-tolerant control strategies are employed.

\section*{Appendix}

\subsection{Proof of Lemma 1}

In order to make this Note self-contained, the following lemma is introduced:

\textbf{Lemma 3 }\cite{Goodwin}. If $\Omega$ is a nonempty convex set in $%
\mathbb{R}
^{4}$ and $F_{0}$ is not an interior point of $\Omega$, then there is a
nonzero vector $k$ such that $k^{T}\left(  F-F_{0}\right)  \leq0$ for each
$F\in cl\left(  \Omega\right)  $, where $cl\left(  \Omega\right)  $ is the
closure of $\Omega$.

Then according to \emph{Lemma 3},

(i)$\Rightarrow$(ii): Suppose that (i) holds. It is easy to see that all the
eigenvalues of $A^{T}$ are zero. By solving the linear equation $A^{T}v=0$,
all the eigenvectors of $A^{T}$ are expressed in the following form%
\begin{equation}
v=[0\text{ }0\text{ }0\text{ }0\text{ }k_{1}\text{ }k_{2}\text{ }k_{3}\text{
}k_{4}]^{T} \label{v}%
\end{equation}
where $v\neq0,k=[k_{1}$ $k_{2}$ $k_{3}$ $k_{4}]^{T}\in%
\mathbb{R}
^{4}$, and $k\neq0$. With it,%
\begin{equation}
v^{T}Bu=-k_{1}\frac{T-m_{a}g}{m_{a}}+k_{2}\frac{L}{J_{x}}+k_{3}\frac{M}{J_{y}%
}+k_{4}\frac{N}{J_{z}}. \label{vBftau}%
\end{equation}
By \emph{Lemma 3}, if $G$ is not an interior point of $\Omega$, then $u=0$ is
not an interior point of $\mathcal{U}$. Then, there is a nonzero
$k_{u}=[k_{u1}$ $k_{u2}$ $k_{u3}$ $k_{u4}]^{T}$ satisfying
\[
k_{u}^{T}u=k_{u1}\left(  T-m_{a}g\right)  +k_{u2}L+k_{u3}M+k_{u4}N\leq0
\]
for all $u\in\mathcal{U}$. Let
\begin{equation}
k=[-k_{u1}m_{a}\text{ }k_{u2}J_{x}\text{ }k_{u3}J_{y}\text{ }k_{u4}J_{z}]^{T}
\label{k}%
\end{equation}
then $v^{T}Bu\leq0$ for all $u\in\mathcal{U}$ according to (\ref{vBftau}),
which contradicts \emph{Theorem 1}.

(ii)$\Rightarrow$(i): As all the eigenvectors of $A^{T}$ are expressed in the
form expressed by equation (\ref{v}), then%
\[
v^{T}Bu=k^{T}J_{f}^{-1}u
\]
according to equation (\ref{linear}) and (\ref{v}) where $k\neq0$. Then there
is no nonzero $v\in%
\mathbb{R}
^{8}$ expressed by (\ref{v}) satisfying $v^{T}Bu\leq0$ for all $u\in
\mathcal{U}$ is equivalent to that there is no nonzero $k\in%
\mathbb{R}
^{4}$ satisfying $k^{T}J_{f}^{-1}u\leq0$ for all $u\in\mathcal{U}$. Supposing
that (ii) is valid, then $u=0$ is an interior point of $\mathcal{U}$. There is
a neighbourhood $\mathcal{B}\left(  0,u_{r}\right)  $ of $u=0$ belonging to
$\mathcal{U}$, where $u_{r}>0$ is small and constant. (ii)$\Rightarrow$(i)
will be proved by counterexamples.

Supposing that condition (i) does not hold, then there is a $k\neq0$
satisfying $k^{T}J_{f}^{-1}u\leq0$ for all $u\in\mathcal{U}$. Without loss of
generality, let $k=\left[  k_{1}\text{ }\ast\text{ }\ast\text{ }\ast\right]
^{T}$ where $k_{1}\neq0$ and $\ast$ indicates an arbitrary real number. Let
$u_{1}=\left[  \varepsilon\text{ }0\text{ }0\text{ }0\right]  ^{T}$ and
$u_{2}=\left[  -\varepsilon\text{ }0\text{ }0\text{ }0\right]  ^{T}$ where
$\varepsilon>0$; then $u_{1},u_{2}\in\mathcal{B}\left(  0,u_{r}\right)  $ if
$\varepsilon$ is sufficiently small. As $k^{T}J_{f}^{-1}u\leq0$ for all
$u\in\mathcal{B}\left(  0,u_{r}\right)  $, then $k^{T}J_{f}^{-1}u_{1}\leq0$
and $k^{T}J_{f}^{-1}u_{2}\leq0$. According to equation (\ref{linear}),%
\[
-\frac{k_{1}\varepsilon}{m_{a}}\leq0,\frac{k_{1}\varepsilon}{m_{a}}\leq0.
\]
This implies that $k_{1}=0$ which contradicts the fact that $k_{1}\neq0$.

Then, condition (i) holds.

(ii)$\Leftrightarrow$(iii): According to the definition of $\rho\left(
G,\partial\Omega\right)  $, if $\rho\left(  G,\partial\Omega\right)  \leq0$,
then $G$ is not in the interior of $\Omega$, and if $\rho\left(
G,\partial\Omega\right)  >0$, then $G$ is an interior point of $\Omega$.

This completes the proof.

\subsection{Proof of Theorem 3}

\emph{Theorem 3} will be proved in the following 3 steps.

\emph{Step 1. Obtain the equations (\ref{equ_hyper_bound}), which are the
projection of parallel boundaries in }$\mathcal{F}$\emph{ by the map }$B_{f}%
$\emph{.}

The results in \cite{Degree} are referred to in order to complete this step.
First, (\ref{F=Bff}) is rearranged as follows:%

\begin{equation}
F=\left[
\begin{array}
[c]{cc}%
B_{1,j} & B_{2,j}%
\end{array}
\right]  \left[
\begin{array}
[c]{c}%
f_{1,j}\\
f_{2,j}%
\end{array}
\right]  \label{tau_map}%
\end{equation}
where $f_{1,j}=[f_{S_{1}\left(  j,1\right)  }$ $f_{S_{1}\left(  j,2\right)  }$
$f_{S_{1}\left(  j,3\right)  }]^{T}\in%
\mathbb{R}
^{3}$, $f_{2,j}=[f_{S_{2}\left(  j,1\right)  }$ $\cdots$ $f_{S_{2}\left(
j,m-3\right)  }]^{T}\in%
\mathbb{R}
^{m-3},$ $j=1,\cdots,s_{m}$. Write (\ref{tau_map}) more simply as%
\begin{equation}
F=B_{1,j}f_{1,j}+B_{2,j}f_{2,j} \label{tau_map_simple}%
\end{equation}
If the rank of $B_{1,j}$ is 3, there exists a 4 dimensional vector $\xi_{j}$
such that%
\[
\xi_{j}^{T}B_{1,j}=0,\left\Vert \xi_{j}\right\Vert =1.
\]
Therefore, multiplying $\xi_{j}^{T}$ on both sides of (\ref{tau_map_simple})
results in%
\begin{equation}
\xi_{j}^{T}F-\xi_{j}^{T}B_{2,j}f_{2,j}=0. \label{equ_hyper}%
\end{equation}

According to \cite{Degree}, $\partial\Omega$ is a set of hyperplane segments,
and each hyperplane segment in $\partial\Omega$ is the projection of a 3
dimensional boundary hyperplane segment of $\mathcal{F}$. Each 3 dimensional
boundary of the hypercube $\mathcal{F}$ can be characterized by fixing the
values of $f_{2,j}$ at the boundary value, denoted by $\bar{f}_{2,j}$, where
\begin{equation}
\bar{f}_{2,j}\in\Pi_{i=1}^{m-3}\left\{  0,K_{S_{2}\left(  j,i\right)
}\right\}  \label{f2ji}%
\end{equation}
and allowing the values of $f_{1,j}$ to vary between their limits given by
$\mathcal{F}$, where $f_{1,j}\in\Pi_{i=1}^{3}\left[  0,K_{S_{1}\left(
j,i\right)  }\right]  $. Then for each $j$, if rank $B_{1,j}=3$, a group of
parallel hyperplane segments $\Gamma_{\Omega,j}=\left\{  l_{\Omega
,j,k},k=1,\cdots,2^{m-3}\right\}  $ in $\Omega$ is obtained, and each
$l_{\Omega,j,k}$ is expressed by%
\begin{equation}
l_{\Omega,j,k}=\left\{  X|\xi_{j}^{T}X-\xi_{j}^{T}B_{2,j}\bar{f}_{2,j}%
=0,X\in\Omega,\bar{f}_{2,j}\in\Pi_{i=1}^{m-3}\left\{  0,K_{S_{2}\left(
j,i\right)  }\right\}  \right\}  \label{equ_hyper_bound}%
\end{equation}
where $\xi_{j}$ is the normal vector of the hyperplane segments.

\emph{Step 2. Compute the distances from the center }$F_{c}$\emph{ to all the
elements of }$\partial\Omega$\emph{.}

It is pointed out that, not all the hyperplane segments in $\Gamma_{\Omega,j}$
specified by equations (\ref{equ_hyper_bound}) belong to $\partial\Omega$. In
fact, for each $j$, only two hyperplane segments specified by equations
(\ref{equ_hyper_bound}) belong to $\partial\Omega$, denoted by $\Gamma
_{\Omega,j,1}$ and $\Gamma_{\Omega,j,2}$, $j\in\left\{  1,\cdots
,s_{m}\right\}  $, which are symmetric about the center $F_{c}$ of $\Omega$.
The center of $\mathcal{F}$ is $f_{c}$, then $F_{c}$ is the projection of
\thinspace$f_{c}$ through the map $B_{f}$ and is expressed as follows%
\begin{equation}
F_{c}=B_{f}f_{c} \label{tau_c}%
\end{equation}
where $f_{c}=\frac{1}{2}[K_{1}$ $K_{2}\ \cdots$ $K_{m}]^{T}\in%
\mathbb{R}
^{m}$. Then the distances from $F_{c}$ to the hyperplane segments given by
(\ref{equ_hyper_bound}) are computed by%
\begin{align}
d_{\Omega,j,k}  &  =\left\vert \xi_{j}^{T}F_{c}-\xi_{j}^{T}B_{2,j}\bar
{f}_{2,j}\right\vert \nonumber\\
&  =\left\vert \xi_{j}^{T}B_{2,j}\left(  \bar{f}_{2,j}-f_{c,2}\right)
\right\vert \nonumber\\
&  =\left\vert \xi_{j}^{T}B_{2,j}\bar{z}_{j}\right\vert \label{d}%
\end{align}
where $k=1,\cdots,2^{m-3},$ $f_{c,2}=\frac{1}{2}[K_{S_{2}\left(  j,1\right)
}$ $K_{S_{2}\left(  j,2\right)  }\ \cdots$ $K_{S_{2}\left(  j,m-3\right)
}]^{T}\in%
\mathbb{R}
^{m-3}$, $\bar{f}_{2,j}$ is specified by (\ref{f2ji}), and $\bar{z}_{j}%
=\bar{f}_{2,j}-f_{c,2}$.

\textbf{Remark 4}. The distances from $F_{c}$ to the hyperplane segments given
by (\ref{equ_hyper_bound}) are defined by $d_{\Omega,j,k}=\min\left\{
\left\Vert X-F_{c}\right\Vert ,X\in l_{\Omega,j,k}\right\}  $, $k=1,\cdots
,2^{m-3}.$

The distances from the\ center $F_{c}$ to $\Gamma_{\Omega,j,1}$ and
$\Gamma_{\Omega,j,2}$ are equal, which is given by%
\begin{equation}
d_{j,\max}=\max\left\{  d_{\Omega,j,k},k=1,\cdots,2^{m-3}\right\}
\label{dj_bar}%
\end{equation}
Since $\bar{z}_{j}\in Z=\frac{1}{2}\Pi_{i=1}^{m-3}\left\{  -K_{S_{2}\left(
j,i\right)  },K_{S_{2}\left(  j,i\right)  }\right\}  ,k=1,\cdots,2^{m-3}$,
\begin{equation}
d_{j,\max}=\frac{1}{2}\text{sign}\left(  \xi_{j}^{T}B_{2,j}\right)
\Lambda_{j}\left(  \xi_{j}^{T}B_{2,j}\right)  ^{T} \label{dj_bar_max}%
\end{equation}
according to (\ref{sign_function}) (\ref{d}) and (\ref{dj_bar}), where
$\Lambda_{j}$ is given by (\ref{Lanmda}).

\emph{Step 3. Compute }$\rho\left(  G,\partial\Omega\right)  $\emph{.}

As $G$ and $F_{c}$ are known, the vector $F_{Gc}=F_{c}-G$ is projected along
the direction $\xi_{j}$ and the projection is given by%
\begin{equation}
d_{Gc}=\xi_{j}^{T}F_{Gc}. \label{lGc}%
\end{equation}
Then if $G\in\Omega$, the minimum of the distances from $G$ to both
$\Gamma_{\Omega,j,1}$ and $\Gamma_{\Omega,j,2}$ is
\begin{equation}
d_{j}=d_{j,\max}-\left\vert d_{Gc}\right\vert \label{d_tau0_p1}%
\end{equation}
But if $G\in\Omega^{C}$, $d_{j}$ specified by (\ref{d_tau0_p1}) may be
negative. So the minimum of the distances from $G$ to both $\Gamma
_{\Omega,j,1}$ and $\Gamma_{\Omega,j,2}$ is $\left\vert d_{j}\right\vert $.
According to (\ref{tau_c}) (\ref{dj_bar_max}) (\ref{lGc}) and (\ref{d_tau0_p1}%
),
\[
d_{j}=\frac{1}{2}\text{sign}\left(  \xi_{j}^{T}B_{2,j}\right)  \Lambda
_{j}\left(  \xi_{j}^{T}B_{2,j}\right)  ^{T}-\left\vert \xi_{j}^{T}\left(
B_{f}f_{c}-G\right)  \right\vert ,j=1,\cdots,s_{m}.
\]
But if rank $B_{1,j}<3$, the 3 dimensional hyperplane segments are planes,
lines, or points in $\partial\Omega$ or $\Omega$ and $\left\vert
d_{j}\right\vert $ will never be the minimum in $\left\vert d_{1}\right\vert
$, $\left\vert d_{2}\right\vert $, $\cdots$, $\left\vert d_{s_{m}}\right\vert
$. The distance $d_{j}$ is set to $+\infty$ if rank $B_{1,j}<3$. The purpose
of this is to exclude $d_{j}$ from $\left\vert d_{1}\right\vert $, $\left\vert
d_{2}\right\vert $, $\cdots$, $\left\vert d_{s_{m}}\right\vert $. In practice,
$+\infty$ is replaced by a sufficiently large positive number (for example,
$d_{j}=10^{6}$).\textbf{ }If $\min\left(  d_{1},d_{2},\cdots,d_{s_{m}}\right)
\geq0$, then $G\in\Omega$ and $\rho\left(  G,\partial\Omega\right)
=\min\left(  d_{1},d_{2},\cdots,d_{s_{m}}\right)  .$ But if $\min\left(
d_{1},d_{2},\cdots,d_{s_{m}}\right)  <0$, which implies that at least one of
$d_{j}<0,j\in\left\{  1,\cdots,s_{m}\right\}  $, then $G\in\Omega^{C}$ and
$\rho\left(  G,\partial\Omega\right)  =-\min\left(  \left\vert d_{1}%
\right\vert ,\left\vert d_{2}\right\vert ,\cdots,\left\vert d_{s_{m}%
}\right\vert \right)  $ according to (\ref{Rou}).

Then $\rho\left(  G,\partial\Omega\right)  $ is computed by%
\begin{equation}
\rho\left(  G,\partial\Omega\right)  =\text{sign}\left(  \min\left(
d_{1},d_{2},\cdots,d_{s_{m}}\right)  \right)  \min\left(  \left\vert
d_{1}\right\vert ,\left\vert d_{2}\right\vert ,\cdots,\left\vert d_{s_{m}%
}\right\vert \right)  . \label{d_min}%
\end{equation}
This is consistent with the definition in (\ref{Rou}).

\section{Acknowledgment}

This work is supported by the National Natural Science Foundation of China
(No. 61473012) and the "Young Elite" of High Schools in Beijing City of China
(No. YETP1071).

\end{document}